\documentclass[prb,twocolumn,showpacs]{revtex4}
\usepackage{graphicx}
\usepackage{epsfig}
\usepackage{dcolumn}

\begin{document}

\title{ Tail States below the Thouless Gap in SNS junctions :\\
Classical Fluctuations.}
\author{Alessandro Silva}
\affiliation{Department of Physics and Astronomy, Rutgers University, 136 Frelinghuysen
Road, Piscataway 08854, New Jersey, USA.}
\date{\today }

\begin{abstract}
We study the tails of the density of states (DOS) in a diffusive superconductor-normal metal-superconductor
(SNS) junction below the Thouless gap. We show that long-wave fluctuations of the concentration of
impurities in the normal layer lead to the formation of subgap quasiparticle states, and calculate the
associated subgap DOS in all effective dimensionalities. We compare the resulting tails with those arising
from mesoscopic gap fluctuations, and determine the dimensionless parameters controlling which contribution
dominates the subgap DOS. We observe that the two contributions are formally related to each other by a dimensional
reduction.
\end{abstract}


\maketitle

\section{Introduction.}
\label{sec0}

The properties of hybrid superconductor-normal metal structures (SN) continue
to attract considerable attention both experimentally~\cite{Gueron} and theoretically~\cite{Brouwer, Ossipov,
Golubov, Zhou, Ivanov, Bruder, Brouwer1, Ostrovsky}, though
the fundamental process governing the physics of such systems, Andreev reflection~\cite{Andreev},
has been discovered long ago. In fact, while it is well known that generically the proximity to a superconductor
leads to a modification of the density of states in the normal metal, the nature and
extent of this effect depends on the details the hybrid structure. In particular, it was
recently pointed out~\cite{Brouwer} that
when a closed mesoscopic metallic region is contacted on one side to a superconductor,
the resulting DOS turns out to depend on its shape.
If integrable, the DOS is finite everywhere but at the Fermi level, where it vanishes as
a power law. On the contrary, in a generic chaotic metallic region one expects the opening of
a gap around the Fermi level, the Thouless gap~\cite{Ossipov}.
In analogy with the considerations above, a diffusive metallic region
sandwiched between two bulk superconducting electrodes has been predicted to
have a gapped density of states, the gap being at energies comparable to to the Thouless energy
$E_{Th}=D/L_z^2$, where $D$ is the diffusion constant and $L_z$ the width of the normal
layer~\cite{Golubov, Zhou, Ivanov, Bruder} [see Fig.1].

In a diffusive SNS structure with transparent SN interfaces, the density of states in the normal part,
averaged over its thickness, and at energies $E$ right above the gap edge $E_g \simeq 3.12 E_{Th}$,
is $\nu \propto 1/\pi V\;\sqrt{(E-E_g)/\Delta_0^3}$, where
$\Delta_0=(E_g \delta^2)^{1/3}$, $\delta=1/(\nu_0 V)$, and $V=L_xL_yL_z$ is the volume of the
normal region. This dependence is reminiscent of the density of states at the edge of a Wigner
semicircle in Random Matrix Theory [RMT],  $\Delta_0$ being the effective level
spacing right above the gap edge. Using this analogy, Vavilov et al.~\cite{Brouwer1} realized
that the disorder averaged DOS should not display a real gap, but
have exponentially small tails below the gap edge, analogous to the Tracy-Widom
tails~\cite{Tracy} in RMT. A rigorous study in terms of a Supersymmetric Sigma Model
description of the SNS structure has shown that this is indeed the case~\cite{Ostrovsky}. However,
in analogy to the theory of Lifshits tails~\cite{Lifshits} in disordered conductors,
the nature of the resulting subgap quasiparticle states depends additionally on the effective dimensionality
$d$, determined by comparing the interface length scales $L_x,L_y$, with the typical length scale
of a subgap quasiparticle state, $L_{\bot}$. In particular, if $L_x \gg L_{\bot}>L_y$ or $L_x,L_y \gg L_{\bot}$
the subgap quasiparticle states are localized either in the $x$ direction or in the $x-y$ plane along the interface,
respectively. Correspondingly, the asymptotic tails of the DOS deviate from the universal RMT result,
applicable only in the zero dimensional case [$L_x,L_y < L_{\bot}$].

The analogy with RMT applies, within the appropriate symmetry
class, to other physical situations, such as diffusive
superconductors containing magnetic
impurities~\cite{Brouwer1,Lamacraft,Aleiner}, and superconductors
with inhomogeneous coupling constants~\cite{Meyer}. In both cases,
at mean field level the density of states has a square root
singularity close to the gap edge~\cite{Abrikosov,Larkin}.
Correspondingly, accounting for mesoscopic RM-like fluctuation,
the disorder averaged density of states has tails below the gap
edge, with an asymptotics similar to the one calculated in
Ref.[\onlinecite{Ostrovsky}] for SNS structures. On the other
hand, in the case of diffusive superconductors containing magnetic
impurities, it was shown~\cite{Me,Balatsky} that, in addition to
\it mesoscopic fluctuations \rm, subgap quasiparticle states can
form as a result of \it classical fluctuations \rm, i.e. long-wave
fluctuations of the concentration of magnetic impurities
associated to their Poissonian statistics. Similarly, also in
superconductors with inhomogeneous coupling constant long-wave
fluctuations of the coarse grained gap lead to the appearance of
subgap quasiparticle states, and consequently to tails of the
DOS~\cite{Larkin}. Interestingly, in both cases the tails
originating from mesoscopic fluctuations and from classical ones
are formally related by a dimensional reduction~\cite{Me}.

In this paper, we close this set of analogies, studying
the contribution to the subgap tails of the DOS in a diffusive
SNS junction arising from long-wave fluctuations of the concentration of impurities
in the normal layer. Combining the results of this analysis with those obtained by Ostrovsky,
Skvortsov, and Feigel'man~\cite{Ostrovsky}, who considered the subgap tails originating
from mesoscopic fluctuations, we provide a consistent picture of the physics of the subgap
states. In particular, a quantitative comparison of the two contribution shows that
mesoscopic fluctuations dominate in long and dirty junctions, while classical fluctuations
dominate in wider and/or cleaner ones. In analogy with diffusive superconductors with
magnetic impurities, and superconductors with inhomogeneous coupling constants, also
in the present case the two contributions to the subgap tails, arising from mesoscopic
and classical fluctuations, are related by a dimensional reduction.

The rest of the paper is organized as follows: in Sec.II we
present the details of the analysis of the subgap DOS arising from
fluctuations of the concentration of impurities $n_{imp}$ in an
SNS junction. In Sec.III, we compare the two contributions to the
subgap DOS associated to mesoscopic and classical fluctuations. In
Sec.IV, we present our conclusions.

\section{Subgap DOS associated to fluctuations of $n_{imp}$.}
~\label{sec01}

Let us start considering a diffusive metallic layer in between two
superconducting bulk electrodes, a geometry represented
schematically in Fig.1. Assuming $k_F l>>1$, where $l$ is the mean
free path, this system can be described in terms of the
quasiclassical approximation. In particular, at mean field level [
i.e., neglecting both mesoscopic and classical fluctuations ],
neglecting electron-electron interaction, and assuming the
thickness of the metallic layer $L_z>>l$,  one can describe the
SNS structure by the Usadel equation~\cite{Usadel,Kopnin}
\begin{eqnarray}\label{Usadel}
\frac{D}{2}\nabla^2 \theta + i\;E\;\sin[\theta]=0,
\end{eqnarray}
where $D=v_{F}^2 \tau/3$ is the diffusion constant, $E$ is the
energy measured from the Fermi level, assumed to be $\mid E \mid
\ll \Delta$, where $\Delta$ is the gap in the bulk electrodes. The
field $\theta$ is related to the quasiclassical Green's functions
and the anomalous Green's function by the relations $g({\bf
r},E)=\cos[\theta({\bf r},E)]$, $ f({\bf r},E)=i \sin[\theta({\bf
r},E)]$. In addition, assuming the interfaces to be perfectly
transparent, the proximity to the two superconducting regions can
be described by the boundary conditions
$\theta(z=\pm L_z/2)=\pi/2$.
{\begin{figure} \epsfxsize=7cm \centerline{\epsfbox{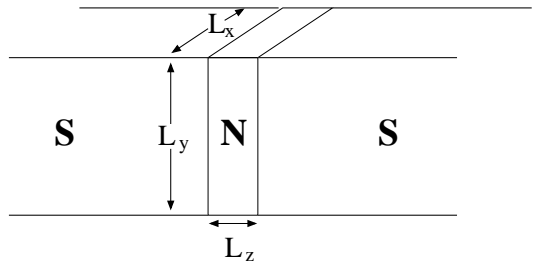}}
\vspace*{0cm} \caption{A schematic plot of an SNS junction: two
bulk superconducting electrodes (S) connected to a diffusive metal
(N) of thickness $L_z$. The interfaces have linear size $L_x$,
$L_y$. } \label{Fig3}
\end{figure}}

It is convenient to measure all lengths in units $L_z$, and set $\theta=\pi/2+i\Psi$. Therefore,
Eq.(\ref{Usadel}) becomes
\begin{eqnarray}\label{UsadelPsi}
\nabla^2 \Psi +2 \frac{E}{E_{Th}} \cosh[\Psi]=0,
\end{eqnarray}
where $E_{Th}=D/L_z^2$ is the Thouless energy. The boundary
conditions for the field $\Psi$ are simply $\Psi(z=\pm 1/2)=0$.

In terms of $\Psi$ the DOS is $\nu=2\nu_0 {\rm Im}[\sinh[\Psi]]$,
where $\nu_0$ is the density of states of the normal metal at the
Fermi level. The DOS can be calculated looking for solutions of
Eq.(\ref{UsadelPsi}) uniform in the $x-y$
plane~\cite{Golubov,Zhou,Ivanov,Ostrovsky}. In particular, for
$E<E_g\equiv C_2 E_{Th}$ [$C_2\simeq 3.122$] all solutions of
Eq.(\ref{UsadelPsi}) are real, implying $\nu=0$. Therefore, one
identifies $E_g$ with the proximity induced gap within the normal
metal layer. The mean field DOS right above $E_g$ averaged over
the $z$ direction is found to be
\begin{eqnarray}\label{resultDOSuniform}
\nu \simeq 3.72\;\nu_0 \sqrt{\frac{E-E_g}{E_g}}.
\end{eqnarray}

Let us proceed analyzing the tails of the DOS at energies $E<E_g$
arising from fluctuations of the concentration of impurities, i.e.
long-wave inhomogeneities in the $x-y$ plane of $1/\tau$. We first
consider an SNS structure such that the linear size of the SN
interfaces is much larger than the thickness of the metallic layer
[$L_x, L_y \gg L_z$]. In the framework of the Usadel description
of the metallic layer [Eq.(\ref{UsadelPsi})] one can account for
long-wave transversal fluctuations of the concentration of
impurities by promoting $E_{Th}$, or equivalently $E_g=C_2
E_{Th}$, to be a position dependent random variable, characterized
by the statistics
\begin{eqnarray}
E_g({\bf x})&=&E_g+\delta E_g({\bf x}),\label{stat1}\\
\langle \delta E_g ({\bf x})  \rangle &=& 0,\label{stat2}\\
\langle \delta  E_g ({\bf x})  \delta E_g({\bf x'}) \rangle
&=& \frac{E_g^2}{n_{d} L_z^d}\;\delta({\bf x}-{\bf x'}),\label{stat3}
\end{eqnarray}
where $d$ is the effective dimensionality of the system,
and $n_{d}$ the effective concentration of impurities.
As shown below, $d$ is determined by comparing the linear sizes of the interface
$L_x,L_y$ to the linear scale of the subgap states  $L_{\bot} \simeq L_z/((E_g-E)/E_g)^{1/4}$.
If $L_x,L_y\gg L_{\bot}$ the system is effectively two dimensional, and $n_{2}=n_{imp} L_z$.
On the other hand, if $L_x < L_{\bot} \ll L_y$ [or $L_y < L_{\bot} \ll L_x$], the system is effectively one dimensional,
and $n_1=n_{imp}\;L_z\;L_x$.

Accounting for these fluctuations, the Usadel equation Eq.(\ref{UsadelPsi}) becomes
\begin{eqnarray}\label{Usadelfluctuations}
\partial_z^2\Psi+\nabla^2_{{\bf x}}\Psi+2C_2 \frac{E}{E_g}(1-\delta \epsilon_g({\bf x}))\cosh[\Psi]=0,
\end{eqnarray}
where $\delta \epsilon_g=\delta E_g/E_g$.

Our purpose is to calculate the DOS averaged over fluctuations of
$\delta E_g$ at energies $E < E_g$. For this sake, let us
introduce $\delta E=E_g-E$, and $\delta\Psi(z,{\bf x})=\Psi(z,{\bf
x})-\Psi_0(z)$, where $\Psi_0$ is the solution of
Eq.(\ref{UsadelPsi}) at $E=E_g$. Expanding
Eq.(\ref{Usadelfluctuations}) and keeping the lowest order
nonlinearity in $\delta\Psi$ one obtains
\begin{eqnarray}\label{intermediate}
(\partial_z^2 + f_0(z)) \delta\Psi+ \nabla^2_{{\bf x}} \delta \Psi +
\frac{g_0(z)}{2} \delta\Psi^2=
g_0(z)(\delta\epsilon-\delta\epsilon_g),
\end{eqnarray}
where $\delta\epsilon=\delta E/E_g$, $g_0(z)= 2C_2
\cosh[\Psi_0(z)]$, and $f_0(z)= 2C_2 \sinh[\Psi_0(z)]$.

In order to simplify further Eq.(\ref{intermediate}), it is useful
to notice that the operator ${\cal H}=-\partial_z^2 -f_0 (z) $,
diagonalized with zero boundary conditions at $\pm 1/2$, admits an
eigenstate $\Phi_0$ with zero eigenvalue. Physically, $\Phi_0$
determines the shape of the mean field $z$-dependent DOS obtained
from Eq.(\ref{UsadelPsi}). Therefore, it is natural to set
\begin{eqnarray}\label{approximation}
\delta \Psi(z,{\bf x}) \simeq \sqrt{A_1/A_2}\;\chi({\bf
x})\;\Phi_0(z),
\end{eqnarray}
with $A_1=\int\;dz\;g_0\;\Phi_0 \simeq 7.18$, and
$A_2=\int\;dz\;\frac{g_0}{2}\;\Phi_0^3 \simeq 2.74$.

Substituting Eq.(\ref{approximation}) in Eq.(\ref{intermediate}),
and projecting the resulting equation on $\Phi_0$, one obtains
\begin{eqnarray}\label{optimal}
\nabla^2 \chi + \chi^2= \delta \epsilon-\delta \epsilon_g({\bf x})
\end{eqnarray}
where we rescaled the length by $(A_1\;A_2)^{-1/4}$, and
\begin{eqnarray}
\langle \delta \epsilon_g({\bf x}) \delta\epsilon_g ({\bf x'})
\rangle= \eta\;\delta ({\bf x}-{\bf x'}),
\end{eqnarray}
with $\eta \equiv (A_1 A_2)^{1/4}/(n_{d}\;L_z^d)$.

Let us now split $\chi=-u+iv$, and obtain the system
\begin{eqnarray}~\label{potential}
&& -\nabla^2 u+ u^2-v^2=\delta\epsilon-\delta\epsilon_g, \\
&& -\frac{1}{2}\nabla^2\;v +u\;v=0.~\label{wavefunction}
\end{eqnarray}
Interestingly, this set of equations is analogous to the equations obtained by
Larkin and Ovchinikov in the context of the study of gap smearing in
inhomogeneous superconductors~\cite{Larkin}, and to the equations obtained by the author and
Ioffe in the context of the study of subgap tails in diffusive superconductors containing
magnetic impurities~\cite{Me}.

Let us now proceed with the calculation of the DOS.
In the present notation, the DOS averaged over the thickness of the normal layer is given by
\begin{eqnarray}\label{DOS2}
\frac{\nu({\bf x},\delta \epsilon \mid \delta\epsilon_g({\bf
x}))}{\nu_0} \simeq 3.72\; v({\bf x},\delta \epsilon \mid
\delta\epsilon_g({\bf x})).
\end{eqnarray}
We are interested in calculating the average density of states $\langle \nu \rangle/\nu_0 \simeq 3.72 \langle v \rangle$
at energies below the Thouless gap [$\delta \epsilon>0$].
In this parameter range, the corresponding functional integral
\begin{eqnarray}\label{functional}
\langle v \rangle \simeq \frac{\int\;D[\delta \epsilon_g] v({\bf
x},\delta \epsilon \mid \delta\epsilon_g({\bf x}))
e^{-1/(2\eta)\;\int d{\bf x} (\delta \epsilon_g({\bf x}))^2}}
{\int\;D[\delta \epsilon_g]
 e^{-1/(2\eta)\;\int d{\bf x} (\delta \epsilon_g({\bf x}))^2}},
\end{eqnarray}
receives its most important contributions by exponentially rare
instanton configurations of $\delta \epsilon_g$ such that, at
specific locations along the interfaces of the junction, $\delta
\epsilon_g({\bf x}) \geq \delta \epsilon$. The remaining task is
to select among all these fluctuations the one that dominates the
functional integral Eq.(\ref{functional}), i.e. the \it optimal
fluctuation \rm.

The action associated to a configuration of $\delta\epsilon_g$ is
\begin{eqnarray}
S=\frac{1}{2\eta} \int\;d{\bf x} (\delta \epsilon_g)^2 \simeq
\int\;d{\bf x} (\nabla^2 u- u^2+\delta\epsilon)^2,
\end{eqnarray}
where we used Eq.(\ref{potential}) to express $\delta \epsilon_g$
in terms of $u,v$, and, with exponential accuracy, neglected the
term $v^2$ in the action. In order to find the optimal fluctuation
one has to find a nontrivial saddle point $u_0$ of $S$, tending
asymptotically to the solution of the homogeneous problem [$u_0
\rightarrow \sqrt{\delta \epsilon}$], and
subject to the constraint of having nontrivial solutions for $v$
of Eq.(\ref{wavefunction}).

Since the normal metal layer is diffusive, and momentum scattering isotropic,
it is natural to assume the optimal fluctuation to be spherically symmetric.
The Euler-Lagrange equation associated to $S$ is
\begin{eqnarray}\label{Euler}
(-\frac{1}{2}\Delta^{(d)}+u)\;(\Delta^{(d)}u-u^2+\delta\epsilon)=0
\end{eqnarray}
where
\begin{eqnarray}
\Delta^{(d)}\equiv\partial_r^2+\frac{d-1}{r}\;\partial_r,
\end{eqnarray}
is the radial
part of the Laplacian in spherical coordinates. An obvious solution to Eq.(\ref{Euler})
is obtained setting
\begin{eqnarray}\label{trivial}
\Delta^{(d)}u-u^2+\delta\epsilon=0.
\end{eqnarray}
This equation is equivalent to the homogeneous Usadel equation
with uniform $E_g$, i.e. Eq.(\ref{optimal}) with
$\delta\epsilon_g=0$.  Though this equation has definitely
nontrivial instanton solutions for $u$ with the appropriate
asymptotics, it is possible to show that the constraint of
Eq.(\ref{wavefunction}) is satisfied only by $v=0$. This is
physically obvious since Eq.(\ref{trivial}) describes a uniform
system where all long-wave fluctuations of $1/\tau$ have been
suppressed, and thus, within the present approximation scheme, the
subgap DOS must vanish. However, it should be pointed out that,
accounting for mesoscopic fluctuations, the instanton solutions of
Eq.(\ref{trivial}) describe the optimal fluctuation associated to
mesoscopic gap fluctuations, as shown in
Ref.[\onlinecite{Ostrovsky}].

Let us now look for the nontrivial saddle point.
Equation (\ref{Euler})
is equivalent to the system
\begin{eqnarray}\label{one}
&&(-\frac{1}{2}\Delta^{(d)}+u) h=0, \\
&&\Delta^{(d)}u-u^2+\delta\epsilon=h.\label{two}
\end{eqnarray}
which can be reduced to a single second order instanton equation setting
$h=(2\partial_r u)/r$. With this substitution, Eq.(\ref{one}) becomes the derivative
of Eq.(\ref{two}), which now reads
\begin{eqnarray}\label{dimred}
\Delta^{(d-2)}u-u^2+\delta\epsilon=0.
\end{eqnarray}
Notice that this equation is, upon reduction of the dimensionality
by $2$, identical in form to the one associated to mesoscopic
fluctuations, Eq.(\ref{trivial}). As we will see later, this
reduction of dimensionality relates in a similar way the
dependence of the action associated to classical and mesoscopic
fluctuations on $\delta \epsilon$.

It is now straightforward to see that the instanton solution $u_0$
of this equation with the appropriate asymptotics describes indeed
the optimal fluctuation, the constraint of Eq.(\ref{wavefunction})
being  automatically satisfied in virtue of Eq.(\ref{one}), with
$v_0 \propto (2\partial_r u_0)/r$. Moreover, the corresponding
optimal fluctuation of $\delta \epsilon_g$ is
$\delta \epsilon_g = 2 \partial_r u_0/r$.

It is clear that the instanton solutions of Eq.(\ref{dimred}) must
have the form $u_0=\sqrt{\delta\epsilon} \upsilon(r/\lambda)$,
with $\lambda=1/(\delta\epsilon)^{1/4}$. The corresponding
equation for $\upsilon(r)$ is $\partial^2_r \upsilon + (d-3)/r
\partial_r \upsilon - \upsilon^2+1=0$. The instanton solution of
this equation can be easily found numerically, and the
corresponding action $S$ calculated. The result is
\begin{eqnarray}
S_d&=& a_d n_d L_z^d\;\delta\epsilon^{\frac{8-d}{4}}
\end{eqnarray}
where the constants $a_d$ are $a_1 \simeq 0.88$, and $a_2 \simeq 7.74$.

Within our approximation scheme, the density of states is $\langle \nu \rangle \propto W \exp[-S]$, where
$W$ is a prefactor due to gaussian fluctuations around the instanton saddle point.
The calculation of $W$ can be performed using the standard
technique due to Zittarz and Langer, and is similar to those reported in
Ref.[\onlinecite{Me},\onlinecite{Larkin}]. To leading order in the saddle point approximation,
the final result is
\begin{eqnarray}\label{finalresult}
\frac{\langle \nu \rangle}{\nu_0} \simeq \beta_d \; \sqrt{n_d\;L_z^d}\;\delta\epsilon^{\frac{d(10-d)-12}{8}}\;e^{-S_d},
\end{eqnarray}
where $\beta_1 \approx 0.1$ and $\beta_2 \approx 0.5$.

The result in Eq.(\ref{finalresult}) relies on a saddle point approximation, which is justified provided
$S_d \gg 1$. This translates into the condition
\begin{eqnarray}\label{condition}
\delta \epsilon \gg \left(\frac{1}{a_d n_d L_z^d}\right)^{\frac{4}{8-d}}.
\end{eqnarray}

As mentioned before, the effective dimensionality, and therefore
the asymptotic density of states, is determined by comparing the
linear size of the optimal fluctuation, in dimensionfull units
$L_{\bot} \simeq L_z \lambda=L_z/\delta\epsilon^{1/4}$, to the
linear dimensions of the interfaces $L_x,L_y$. If $L_x,L_y \gg
L_{\bot}$ the asymptotics is effectively two dimensional [$d=2$],
while for $L_y \gg L_{\bot}, L_x \ll L_{\bot}$ the asymptotic DOS
is effectively one dimensional [$d=1$]. Since $L_{\bot}$ increases
as the energy gets closer to the average gap edge, it is clear
that in any finite size system the applicable asymptotics might
exhibit various crossovers, $2{\rm d} \rightarrow 1{\rm d}
\rightarrow 0{\rm d}$, as $\delta\epsilon \rightarrow 0$. In
particular, the tails are zero dimensional when $L_x,L_y <
L_{\bot}$, in which case the asymptotic form of the DOS is
obtained by calculating the integral
\begin{eqnarray}
\frac{\langle \nu \rangle}{\nu_0} &\simeq& 3.72 \int \frac{d(\delta \epsilon_g)}{\sqrt{2\pi\eta_0}}\;
\sqrt{\delta\epsilon_g-\delta\epsilon}\;e^{-\frac{\delta\epsilon_g^2}{2\eta_0}} \nonumber \\
& \approx & \frac{1}{\delta\epsilon^{3/2}}\;e^{-S_0},
\end{eqnarray}
where $\eta_0=1/(n_{imp}V)$ [$V=L_xL_yL_z$] and $S_0=1/(2\eta_0)\delta\epsilon^2$.

\section{Mesoscopic vs. Classical fluctuations.}

In the previous section we have discussed the asymptotic density
of states below the Thouless gap originating from classical
fluctuations, i.e. inhomogeneities in the concentration of
impurities or equivalently in $1/\tau$. As discussed in the
introduction, this mechanism to generate subgap states is
complementary to mesoscopic fluctuations of the gap edge.

The tails
associated to mesoscopic gap fluctuations have been calculated by Ostrovsky, Feigel'man and
Skvortsov in Ref.[\onlinecite{Ostrovsky}]. To exponential accuracy, the subgap DOS associated to mesoscopic fluctuations
is $\langle \nu \rangle/\nu_0 \propto \exp[-\tilde{S}_d] $, where
\begin{eqnarray}\label{meso}
\tilde{S}_d &\simeq& \tilde{a}_d\; G_d\;
(\delta\epsilon)^{\frac{6-d}{2}},
\end{eqnarray}
where $\tilde{a}_d$ is a constant [$\tilde{a}_0\simeq 1.9$, $\tilde{a}_1\simeq 4.7$, and $\tilde{a}_2 \simeq 10$],
and $G_d$ is the effective dimensionless conductance
\begin{eqnarray}
G_0&=&4\pi \nu_0 D \frac{L_x L_y}{L_z},\\
G_1&=&4\pi \nu_0 D L_x, \\
G_2&=&4\pi \nu_0 D L_z.
\end{eqnarray}
The scale of the optimal fluctuation associated to mesoscopic fluctuations is also
$L_{\bot} \simeq L_z/(\delta\epsilon)^{1/4}$. Therefore, the effective dimensionality
$d$ is to be determined according to the criteria presented in the previous section.

Before discussing the comparison of mesoscopic and classical
fluctuations, let us first explain the rationale behind the
separation these two contributions. Though it is clear that the
only physical fluctuations in a real sample are associated to
fluctuations in the positions of impurities, these fluctuations
can affect the DOS in two ways: \it i)- \rm depress the Thouless
gap edge by increasing locally the scattering rate [classical
fluctuations], or \it ii)- \rm take advantage of interference
effects in the quasiparticle wave functions to generate
quasiparticle states that couple inefficiently to the
superconducting banks [mesoscopic fluctuations]. It makes sense to
think of two types of effects separately if the actions associated
to them are very different in magnitude [$\tilde{S} \gg S$ or vice
versa]. Obviously, in the crossover region, where $S \approx
\tilde{S}$ the separation of these two mechanisms is meaningless,
because the system can take advantage of both at the same time.

With this caveat, let us proceed in the comparison of these two contributions,
starting with the zero dimensional case. Since the dimensionless conductance
is $G_0 \approx E_g/\delta$, where $\delta \approx 1/(\nu_0 V)$
is the level spacing, then the $d=0$ action associated to mesoscopic
fluctuations can be written as
\begin{eqnarray}\label{universal}
\tilde{S}_0 \approx \left(\frac{\delta E}{\Delta_0}\right)^{3/2},
\end{eqnarray}
where $\Delta_0=(E_g \delta^2)^{1/3}$, where $\delta=1/(\nu_0 V)$ is the
level spacing in the metallic layer. Physically, $\Delta_0$ can be
interpreted as being the \it effective \rm
level spacing right above the gap edge. Indeed, from Eq.(\ref{resultDOSuniform})
one sees that
\begin{eqnarray}
\nu \approx \frac{1}{\pi V}\;\sqrt{\frac{\delta E}{\Delta_0^3}}.
\end{eqnarray}
Therefore, the result of Eq.(\ref{universal}) indicates that tails originating
from mesoscopic fluctuations of the gap edge are universal [in $d=0$], in
accordance to the conjecture formulated in Ref.[\onlinecite{Brouwer1}] on the basis
of Random Matrix Theory.
In turn, in the zero dimensional case the action associated to classical fluctuations is
\begin{eqnarray}
S_0 \approx \left(\frac{\delta E}{\delta E_0}\right)^{2},
\end{eqnarray}
where $\delta E_0=E_g/\sqrt{n_{imp} V}$ is the scale of typical fluctuations
of the gap edge associated to fluctuations of the concentration of impurities.
The dimensionless parameter controlling which which mechanism dominates
is therefore
\begin{eqnarray}
\gamma_0=\frac{\Delta_0}{\delta E_0}.
\end{eqnarray}
Clearly, for $\gamma_0 \gg 1$ mesoscopic fluctuations dominate the subgap tails,
while for $\gamma_0 \ll 1$ classical fluctuations give the largest contribution
to the subgap DOS~\cite{detailedcomparison}.

Let us now write $\gamma_0$ in terms of elementary length scales, one can estimate
\begin{eqnarray}\label{gamma0}
\gamma_0 &\approx& \frac{1}{k_F l}\;\frac{1}{\sqrt{k_F^2 \sigma}}
\frac{(L_z/l)^{7/6}}{(L_x L_y/l^2)^{1/6}} \nonumber \\
&\approx& \frac{1}{k_F l}\;
\frac{(L_z/l)^{7/6}}{(L_x L_y/l^2)^{1/6}},
\end{eqnarray}
where we used the fact that the scattering cross section of a single impurity
$\sigma$ is typically of the same order of $\lambda_F^2$. Within the assumptions
of the theory, $\gamma_0$ is the ratio of two large numbers, and therefore
its precise value depends on the system parameters. However, from Eq.(\ref{gamma0})
we see that making the junction longer and longer, i.e. increasing $L_z$,
tends to favor mesoscopic fluctuations. Intuitively, this is due to
the fact that as $L_z$ increases, the dimensionless conductance of the junction
diminishes while the average number of impurities
increases, therefore suppressing the associated fluctuations of the gap edge.
At the same time, increasing the area of the junction, or making them cleaner,
reverses the situation. In summary, mesoscopic fluctuations are favored
in \it long and dirty \rm junctions, while classical fluctuations are favored in
\it wider and/or cleaner \rm ones.

Since in higher dimensionalities the linear scale of the optimal fluctuation
associated to the two mechanism is identical [$L_{\bot}=L_z/(\delta\epsilon)^{1/4}$],
it is possible, and physically suggestive, to reduce the form of the actions
in $d=1,2$ to a zero dimensional action calculated within the typical volume of
the optimal fluctuation. The latter is $V_{\bot}=L_{x}L_{\bot}L_z$ for $d=1$, and
$V_{\bot}=L_{\bot}^2 L_z$ in $d=2$.
For example, for $d=1$ one can write
\begin{eqnarray}
S_1 &\approx& n_{imp}L_x L_{\bot} L_z \;(\delta \epsilon)^2 \nonumber \\
  &\approx& \left(\frac{\delta E}{\delta E_{eff}} \right)^2,
\end{eqnarray}
where $\delta E_{eff}=E_g/\sqrt{n_{imp} V_{\bot}}$. Similarly,
\begin{eqnarray}
\tilde{S}_1 &\approx& \left(\frac{\delta E}{\Delta_{eff}} \right)^2,
\end{eqnarray}
where $\Delta_{eff}=(E_g \delta_{eff}^2)^{1/3}$,
$\delta_{eff}=1/(\nu_0 V_{\bot})$ being the level spacing in the
volume of the optimal fluctuation. In analogy to the zero
dimensional case, one is therefore led to conclude that also in
for one dimensional tails \it long and dirty \rm junctions are
dominated by mesoscopic fluctuations, while \it wider and/or
cleaner \rm junctions favor classical ones. This qualitative
statement is indeed correct, but the proof is complicated by the
energy dependence $L_{\bot}$.

The appropriate way to proceed for $d=1,2$ is to write
the actions associated to classical and mesoscopic fluctuations in compact form as
\begin{eqnarray}
S&=&\left(\frac{E_g-E}{\delta E_d}\right)^{\frac{8-d}{4}},\\
\tilde{S}&=&\left(\frac{E_g-E}{\Delta_d}\right)^{\frac{6-d}{4}}
\end{eqnarray}
where $\delta E_d={E_g}/{(a_d\;n_d L_z^d)^{{4}/{(8-d)}}}$, and
$\Delta_d={E_g}/{(\tilde{a}_d G_d)^{{4}/{(6-d)}}}$. Therefore, the dimensionless
parameter that determines which contributions dominates the subgap DOS is
\begin{eqnarray}
\gamma_d \equiv \frac{\Delta_d}{\delta E_d}.
\end{eqnarray}
If $\gamma_d \gg 1$, the subgap DOS is dominated by mesoscopic gap fluctuations,
and the applicable result is Eq.(\ref{meso}). On the other hand, for $\gamma_d \ll 1$
the DOS below the gap is determined by long-wave fluctuations of $1/\tau$ [Eq.(\ref{finalresult})].
Finally, estimating $\gamma_d$ in terms of elementary length scales, one obtains
\begin{eqnarray}
\gamma_1& \approx & \frac{1}{(k_F l)^{16/35}}\;\frac{(L_z/l)^{8/7}}{(L_x/l)^{8/35}},\\
\gamma_2& \approx & \frac{1}{(k_F l)^{2/3}}\;(L_z/l).
\end{eqnarray}
In analogy to Eq.(\ref{gamma0}), the fact that $\gamma_d$ is proportional to a
power of $L_z/l$ implies that mesoscopic fluctuations are dominant in long junctions,
while the inverse proportionality of $\gamma_d$ on a power of $k_F l$ and of
the linear size of the interface [in $d=0,1$] implies that wide interfaces and/or
cleaner samples may favor the contribution arising from classical fluctuations.

~\label{sec3}

\section{Conclusions}

In this paper, we discussed the effect of inhomogeneous
fluctuations of the concentration of impurities, or equivalently
of $1/\tau$, on the tails of the DOS below the Thouless gap in
diffusive SNS junctions. We have shown that these classical
fluctuations lead to the formation of subgap quasiparticle states
and are complementary to mesoscopic fluctuations in determining
the asymptotic DOS. Finding the dimensionless parameter that
controls which mechanism gives the dominant contribution to the
subgap tails, one finds that, qualitatively,  mesoscopic
fluctuations are favored in long and dirty junctions, while
classical ones dominate in wider and/or cleaner ones.

We have observed that, as for diffusive superconductors containing
magnetic impurities, and for diffusive superconductors with an
inhomogeneous coupling constant, the two contributions are
formally related by a dimensional reduction by $2$, both at the
level of instanton equations determining the optimal fluctuation,
and in the dependence of the DOS on the distance from the gap edge
$\delta\epsilon$. As in other physical systems~\cite{Cardy}, it is
natural to expect that supersymmetry is at the root of dimensional
reduction also in this context. This fact could in principle be
elucidated generalizing the Sigma Model describing mesoscopic
fluctuations to include the physics associated to classical
fluctuations.

\section{Acknowledgements.}

I would like to thank  E. Lebanon, A. Schiller, and especially
L. B. Ioffe and M. M\"{u}ller for discussions. This work is supported by NSF
grant DMR 0210575.


\begin{thebibliography}{99}
\bibitem{note}  Nodes in $v_0$ are not admissible because of the analytic
properties of the quasiclassical Green's functions.
\bibitem{detailedcomparison} This is true even though the exponents
of the energy dependendence of the two actions are different [$S \propto \delta E^{2}$
and $\tilde{S} \propto \delta E^{3/2}$]. Indeed, while for $\gamma_0 \ll 1$ one has $S \ll \tilde{S}$
only for $\delta E < \delta E_0 / \gamma_0^3$, at the crossover point
$S=\tilde{S} \approx 1/\gamma_0^6 >>>1$, meaning that $\nu \propto \exp[-S]$ is already,
for all practical purposes zero.
\end{thebibliography}
\end{document}